\documentclass[pra,10pt,aps,superscriptaddress,twocolumn,amsmath,amssymb,nofootinbib,longbibliography]{revtex4-1}

\usepackage{natbib}
\usepackage[colorlinks]{hyperref}
\hypersetup{linkcolor=blue,citecolor=blue,filecolor=blue,urlcolor=blue}

\usepackage{stackengine}
\stackMath

\usepackage{graphicx}
\usepackage{bm}
\usepackage{color}
\usepackage{soul}

\def\pra{Phys. Rev. A}

\def\prl{Physical Review Letters}

\def\nat{Nature}

\begin{document}

\title{Aspects of wave propagation in a nonlinear medium: birefringence and the second-order magnetoelectric coefficients}

\author{Vitorio A. \surname{De Lorenci}}
\email{delorenci@unifei.edu.br}
\affiliation{Instituto de F\'{\i}sica e Qu\'{\i}mica, Universidade Federal de Itajub\'a, \\
Itajub\'a, Minas Gerais 37500-903, Brazil}

\begin{abstract}
Magnetoelectric materials have the interesting property of exhibiting polarization induced by a magnetic field or magnetization induced by  an electric field. As a consequence, a multitude of effects can be produced by means of controllable external fields. 
A method for deriving phase velocities and its corresponding polarization vectors for light rays in nonlinear optical materials in a nondispersive regime is here revisited and used to study wave propagation in a certain class of second-order magnetoelectric media. 
In particular, the birefringence effect is theoretically examined and it is shown that it can be used as a tool to obtain most of the second-order magnetoelectric coefficients $\beta_{ijk}$ of a material having an isotropic linear sector. Estimates of the effect are presented. Some optical properties of nonlinear materials presenting a natural optic axis are also discussed. 
\end{abstract}

\maketitle

\section{Introduction}
Magnetoelectric materials, which includes chiral media, are characterized by the property of presenting polarization induced by a magnetic field and magnetization induced by an electric field.
First investigations of light propagation in such systems date back to the 19th century, and includes important accounts by, among others,  Arago~\cite{arago}, Biot~\cite{1815Biot}, Pasteur~\cite{1848Pasteur},  Fresnel~\cite{fresnell}, Curie~\cite{1894Curie}, who discussed the  existence of magnetoelectric optical effects by means of simple symmetry assumptions, and Drude \cite{1900leop.book.....D}, who presented the constitutive relations for an isotropic chiral medium.
%
%
Modern theoretical and experimental investigations began to appear around the 1960s \cite{1959OptSp...6...49F,1959OptSp...6..237F,1960PhRv..118.1396P,1960-Dzyaloshinskii,1960-Astrov,1965PMag...11..647F,1967PMag...15..687B}. For instance, the magnetoelectric effect was measured \cite{1960-Astrov} in ${\rm Cr}_2{\rm O}_3$ crystals, confirming a theoretical description of half a year before \cite{1960-Dzyaloshinskii}.
%

%
The study of the physical properties of these materials has become an increasingly active research topic over the years \cite{1973-schmid,1989LNP...335.....L,2009EPJB...71..299R,Lindell1994book,2005JPhD...38R.123F,2021APLM....9e0401H,2021APLM....9d1114L}.
In spite of being tiny effects, the fact that they can be externally controlled by applied fields makes these systems particularly promising for eventual technological applications. Nowadays, with the astonishing improvement in the accuracy of electromagnetic-based apparatus, it seems clear that the conception of new devices working on the submicron scale and operating at extremely low power consumption is at hand \cite{2006Natur.442..759E,2021APLM....9e0401H,2021APLM....9d1114L}.

The list of recent achievements in the domain of magnetoelectric materials is enormous and covers several branches of science (for a review see Ref. \cite{2021APLM....9d1114L} and references therein).
Just to mention a few recent results, we refer to the discussion of nonreciprocal transmission of light \cite{PhysRevLett.123.077401}, the use optical magnetoelectric effect to identify  antiferromagnetic domains  \cite{2018PhRvL.121e7601K}, the proposal of submillimetric magnetoelectric antennas for wireless energy harvesting and magnetic field sensing \cite{2021NatCo..12.3141Z},  the electric field control of natural optical activity \cite{2021Sci...372..496M}, the evidence for a Lorentz force acting on a light ray in multiferroic materials \cite{2005PhRvL..95w7402S}, the usefulness of relativity theory in the understanding the phenomenology of the magnetoelectric effect \cite{2009EPJB...71..321H} (see also Ref.~\cite{1960PhRv..118.1396P}), the study of plane-wave propagation in the scenario of extreme magnetoelectric materials  \cite{2016ITAP...64.5382L},  and the possibility of inducing triple-refringence of light \cite{2012PhRvA..86a3801D,2019PhRvA..88841} in magnetoelectric metamaterials \cite{2014RvMP...86.1093L}, among many others.

The above-mentioned trirefringence is related to the possible existence of three different solutions to the wave equation that propagate in the same direction but with different phase velocities, each one associated with an independent wave polarization vector. In terms of the existence of propagating waves in a natural medium, up to now only birefringence was experimentally observed, and it is worth mentioning that its use in science and technology is of great importance. It should also be mentioned that trirefringence appeared earlier in the context of vector potentials in isotropic birefringent media. It was shown \cite{lakhtakia92} that, even though the electromagnetic fields are birefringent in such systems, three independent solutions for the vector potential are possible.

It is worth briefly presenting the nomenclature used to identify the different manifestations of the magnetoelectric effect (ME). A direct ME refers to an electric polarization induced by a magnetic field, while a converse ME refers to a magnetization induced by an electric field. The effect can be linear or nonlinear, depending on the power of the electric and magnetic fields appearing in the expansion of polarization ($\vec P$) and magnetization ($\vec M$) vectors.
The most general linear ME occurs in optically anisotropic media (and is also referred to as the bianisotropic ME \cite{cheng1968,2010eabf.book.....M}), in which both the direct and converse ME can occur at the same time, and the directions of $\vec P$ and $\vec M$ do not necessarily coincide with the directions of the electric and magnetic fields that induce the effect. Anisotropicity can occur in any or all three possible sectors: dielectric, magnetic, or magnetoelectric. When the medium is such that polarization and magnetization are produced in the same direction of the inducing fields, the medium is denoted as optically isotropic. Hence, considering only the linear sector in the polarization and magnetization expansions, these effects are usually called linear anisotropic (or bianisotropic) and isotropic (or bisotropic) MEs. In the linear sector, the inducing fields can couple to the optical properties of the medium by means of certain coefficients. They are the first-order susceptibilities and are usually described by rank-two tensors which, in components notation, are the well known dielectric $\chi^{{}_{(1)}}_{ij}$, magnetic $\tilde\chi^{{}_{(1)}}_{ij}$, and magnetoelectric  $\alpha_{ij}$ susceptibilities. 
When nonlinear contributions are considered, the effects become appreciably richer, but much smaller in magnitude. The first nonlinear contribution to $\vec P$ or $\vec M$ is quadratic in the electromagnetic fields. Thus, it can appear by means of couplings with the {\em square} of the electric or magnetic field, or even as a cross terms of these fields. Such ME are thus called {\em second-order} effects. To describe the coupling of the optical properties of the medium with the square of the fields, a rank-three coupling tensor, commonly called a second order susceptibility tensor, is needed. As there are four possible couplings, there are four of such tensorial quantities describing the optical properties of the medium. They are the second-order dielectric $\chi^{{}_{(2)}}_{ijk}$, magnetic $\tilde\chi^{{}_{(2)}}_{ijk}$,  and magnetoelectric $\beta_{ijk}$ and $\gamma_{ijk}$ susceptibilities. All these coefficients originate in the expansion of the density of  the free-energy of the medium \cite{landau,2009EPJB...71..299R} in terms of the electromagnetic fields, as will be properly discussed later on in this text.
It may happen that, due to symmetry aspects of the crystalline lattice, some of the nonlinear contributions are absent. For instance, in systems showing inversion symmetry the second-order coupling $\chi^{{}_{(2)}}_{ijk}$ is identically zero. 
A possible way of referring to the above-mentioned effects is by means of the kernel letter associated with each specific coupling. For instance, linear magnetoelectric effects  can generically be referred to as $\alpha$-MEs, which can be isotropic or anisotropic, and similarly the nonlinear effects can be referred to as $\beta$-MEs or $\gamma$-MEs.
Depending on the physical regime assumed in the investigation, these coefficients can be coupled to external static fields or variable fields. In the  latter case they will be dependent on the frequency of the applied field and issues such as frequency conversion may be important. 

 %
 
The main objective of this work is to investigate monochromatic plane-wave propagation in a certain class of nonlinear magnetoelectric materials in a lossless and dispersionless regime (see however the discussion in Appendix~\ref{nonlocal}). 
Basically, the discussion is constrained to a regime in which the electric and magnetic fields can be written as a large constant part (background field) plus a small oscillating contribution (the wave field). The wave fields thus propagate in a medium whose optical nonlinearities are activated by the background fields. The analysis is restricted to the case in which nonlinearities are local in space and time. The discussed methodology can be straightforwardly applied to study wave propagation in many optical systems.
Of particular interest are birefringence effects related to second-order nonlinear magnetoelectric contribution (related to $\beta$-ME) to the polarization and magnetization of the material. 

Following recent investigations \cite{2008PhRvD..78d5015D,*2010PhRvE..82c6605D,2012PhRvA..86a3801D,2019PhRvA..88841}, we start in the next section  by stating Maxwell's field equations in general material media using component notation.  Basic aspects of light propagation in nonlinear materials are summed up in a brief but self-contained presentation in Sec.~\ref{wave-propagation}, where we discuss the eigenvalue problem whose solutions lead to the phase velocities and the corresponding polarization vectors of light rays in such media.  
In particular, these results are used in Sec.~\ref{second-order} to study wave propagation in a second-order nonlinear magnetoelectric medium. It is shown that the eigenvalue problem results in a fourth-degree equation for the phase velocity of light rays propagating in such medium, as expected. The case of a nonlinear medium exhibiting an isotropic linear sector is examined in Sec.~\ref{iso}. It is shown that birefringence occurs in most of the configurations where an external field is present. Furthermore, the ordinary ray is not sensitive to the nonlinearities of the medium. The behavior of some idealized models are numerically examined, including their normal surfaces. The expression for the group velocities are also presented and it is shown that despite the fact that their magnitude coincides with the magnitude of the corresponding phase velocities, the group velocity related to the extraordinary ray propagates in a different direction as compared with the extraordinary phase velocity. Additionally, still in the context of systems presenting an isotropic linear sector,  it is shown in Sec.~\ref{beta} how measurements of the birefringence effect lead to the determination of almost all the components of the second-order magnetoelectric coefficient $\beta_{ijk}$. Estimates for birefringence measurements are discussed in Sec.~\ref{estimates}.  Some aspects of nonlinear systems having a natural optic axis are briefly examined in Sec.~\ref{aniso}. Final remarks and conclusions are presented in Sec.~\ref{final}.
Furthermore, a concise review about the possible magnetoelectric couplings up to second order is presented in Appendix \ref{Polarization}, where the polarization and magnetization vectors are given by means of derivatives of the density of free-energy of the material. Finally, a brief discussion of time dispersion and its implications for the approximative method implemented for studying wave propagation is presented in Appendix~\ref{nonlocal}.

To work with more compact results, the three-dimensional component notation for vectorial and tensorial quantities will be mostly used.  Thus, throughout the text Latin indices $i,j,k...$ run from 1 to 3 (the three spatial directions). The Einstein convention for sum is used, which means that  repeated indices in a monomial indicate summation. Partial derivative with respect to coordinate $x_i$ is denoted by $\partial_i$, while the time derivative is $\partial_t$.  The three-dimensional Levi-Civita symbol $\epsilon_{ijk}$ is a completely antisymmetric quantity defined by $\epsilon_{123}=1$. In this notation, the divergence of an arbitrary vector field $\vec s$ is written as $\partial_i s_i$ while the $k$-th component of its curl is written as $\epsilon_{ijk}\partial_i s_j$. The elements of the three-dimensional identity matrix,  ${\rm diag} (1,1,1)$, are represented by $\delta_{ij}$ (the Kronecker delta), which coincides with the metric of the tri-space, $\eta_{ij}$. 
It is helpful to set here the main notation for the wave vector $\vec q$ which, in terms of the Cartesian unit basis vectors $\hat x_i$, reads $\vec q  = q_i \hat x_i$. This vector can also be expressed in terms of its dimensionless unit vector $\hat\kappa$ as $\vec q = q \, \hat \kappa$. In component notation it reads $q_i = q \kappa_i$, such that $q_i q_i = q^2$ and $\kappa_i \kappa_i = 1$.

\section{Preliminaries}

\subsection{Field equations and constitutive relations}
\label{field-equations}
The general form of Maxwell's equations can be presented in component notation as follows,
\begin{eqnarray}
\partial_{i}E_{i}&=&\frac{\rho}{\varepsilon_0},  \hspace{5mm} \epsilon_{ijk}\partial_{i}E_{j}=-\partial_{t}B_{k},
\nonumber 
\\
\partial_{i}B_{i}&=&0,  \hspace{7.3mm} \epsilon_{ijk}\partial_{i}B_{j}= \mu_0\varepsilon_0\partial_{t}E_{k} + \mu_0 J_k \, ,
\nonumber 
\end{eqnarray}
where $\rho$ and $J_k$ describe  total charge and current densities of the system, respectively. These quantities can be written in terms of free charge density $\rho^{\scriptscriptstyle \tt(free)}$ and current density $J^{\scriptscriptstyle \tt(free)} _{k}$ contributions, polarization $P_i$, and magnetization $M_i$,  such that $\rho= \rho^{\scriptscriptstyle \tt(free)} - \partial_{i}P_{i}$ and $J_{k}= J^{\scriptscriptstyle \tt(free)} _{k} + \epsilon_{ijk}\partial_{i}M _{j} + \partial_{t}P_{k}$.
Then, defining the auxiliary fields $D_i \doteq \varepsilon_0 E_i +  P_i$  and $H_i \doteq  B_i / \mu_0  - M_i$, Maxwell's equations read,
\begin{eqnarray}
\partial_{i}D_{i}&=&\rho^{\scriptscriptstyle \tt(free)},  \hspace{4mm} \epsilon_{ijk}\partial_{i}E_{j}=-\partial_{t}B_{k},
\label{1b}
\\
\partial_{i}B_{i}&=&0,  \hspace{9.5mm} \epsilon_{ijk}\partial_{i}H_{j}=\partial_{t}D_{k} + J^{\scriptscriptstyle \tt(free)} _{k}.
\label{2b}
\end{eqnarray}

Depending on the optical properties of a material medium, $P_i$ and $M_i$ can be expanded in terms of the fundamental fields $E_i$ and $B_i$, as discussed in Appendix~\ref{Polarization}. Such relationship leads to the constitutive relations, which can be conveniently presented as,
\begin{eqnarray}
D_{i}&=&\varepsilon_{ij}(\vec E, \vec B) E_{j}+\tilde{\varepsilon}_{ij}(\vec E, \vec B) B_{j},
\label{constitutive1}
\\
 H_{i}&=&\mu_{ij}^{\mbox{\tiny (-1)}}(\vec E, \vec B) B_{j}+\tilde{\mu}_{ij}^{\mbox{\tiny (-1)}}(\vec E, \vec B) E_{j},
\label{constitutive2}
\end{eqnarray}
where the  $\varepsilon_{ij}$, $\tilde{\varepsilon}_{ij}$, $\mu_{ij}^{\mbox{\tiny (-1)}} $, and $\tilde{\mu}_{ij}^{\mbox{\tiny (-1)}} $ 
describe the optical properties of the material and its relationship with external applied fields. They are just auxiliary quantities, in the sense that they are determined by the polarization and magnetization vectors for each material medium in consideration. 
Furthermore, as indicated, they are tensorial functions that generally depend on applied electric and magnetic fields, as occurs in nonlinear systems, and can also depend on thermodynamic variables, such as temperature. For instance, the Pockels electro-optic effect  is associated with an effective dielectric permittivity of the type $\varepsilon_{ij} =\varepsilon_0(\delta_{ij} + \chi^{{}_{(1)}}_{ij} + \chi^{{}_{(2)}}_{ijk}E_k)$, which makes $D_i$ a quadratic function of the applied electric field (see Appendixes~\ref{Polarization} and \ref{nonlocal} for more details about the origin of the susceptibility coefficients in $\varepsilon_{ij}$). 
%
On the other hand, a linear magnetoelectric medium would present all coefficients in Eqs.~(\ref{constitutive1}) and (\ref{constitutive2}) as constants, resulting in a linear dependence between the auxiliary fields $D_i$ and $H_i$ with the electromagnetic fields $E_i$ and $B_i$. In such a particular domain, an extreme magnetoelectric medium  is reported \cite{2016ITAP...64.5382L} to be characterized by $\varepsilon_{ij} =0$ and $\mu_{ij}^{\mbox{\tiny (-1)}}=0$.

It is worth clarifying the notation $\mu_{ij}^{\mbox{\tiny (-1)}}$. In the particular case of a linear and isotropic magnetic medium presenting constant magnetic permeability $\mu$, it  follows that  $\mu_{ij}^{\mbox{\tiny (-1)}} = (1/\mu)\delta_{ij}$ and $\tilde\mu_{ij}^{\mbox{\tiny (-1)}} = 0$, so that Eq.~(\ref{constitutive2}) reduces to $H_i = (1/\mu)B_i$. Hence, $\mu_{ij}^{\mbox{\tiny (-1)}}$ has dimension of inverse of magnetic permeability. On the other hand, the upper symbol ${}^{\mbox{\tiny (-1)}}$ was kept in $\tilde\mu_{ij}^{\mbox{\tiny (-1)}}$ just for sake of symmetry in the notation. 
Additionally, it should be mentioned that there are equivalent ways of expressing the above constitutive relations. For instance we could proceed by keeping only $\varepsilon_{ij}(\vec E, \vec B)$ and $\mu_{ij}^{\mbox{\tiny (-1)}}(\vec E, \vec B)$ in these relations, without loss of generality. However it is mathematically opportune to express them in the above more expanded form.

In a more general scenario, when dispersion is considered, the coefficients in Eqs.~(\ref{constitutive1}) and (\ref{constitutive2}) should be understood as integral operators, so that these relations also describe the causal relationship between the applied fields and the consequential medium response. Some basic aspects about this issue, are discussed in Appendix~\ref{nonlocal}, where a particular example based on a nonlinear medium is discussed. 

\subsection{The wave propagation  in nonlinear media}
\label{wave-propagation}

In what follows, assuming the absence of free charges and currents ($\rho^{\scriptscriptstyle \tt(free)}=0, \; J^{\scriptscriptstyle \tt(free)} _{k}=0$), we investigate the main aspects of light propagation in nonlinear optical material in a lossless nondispersive regime. 

Let the total electric field be split as a sum of a strong  constant background field $E_i^{0}$ plus a weak but rapidly varying wave field $E_i^{\omega}= E_i^{\omega}(\vec r,t)$ contribution as $E_i = E_i^{0} + E_i^{\omega}$, such that  $E_i^{{\omega}}$ can be neglected when compared with $E_i^{0}$, but $\partial_j E_i = \partial_j E_i^{{\omega}}$ and $\partial_{t}E_i = \partial_{t} E_i^{{\omega}}$. An analogous prescription applies to the magnetic field, $B_i = B_i^{0} + B_i^{\omega}$. 
In other words, we are assuming the existence of constant and strong background fields ($E_i^{0}$ and $B_i^{0}$) that induce polarization and magnetization of the medium. The additional contributions, the wave-fields $E^\omega_i$ and $B^\omega_i$, are considered as small perturbations that propagate through the modified medium without inducing appreciable effects on it. This means that second-order terms in the wave field are being neglected. In this approximative regime, issues related to causal behavior of the medium (due to the finiteness of light velocity, there will always be a delay between the applied fields and the medium response) would only be important during the interval of time when the background fields are being turned-on or -off. 

We now look for monochromatic plane-wave solutions of the type $E_j^{\omega} = e_j \,\mbox{exp}[i(\omega t-q_n x_n)]$ and $B_j^{\omega} = b_j \,\mbox{exp}[i(\omega t-q_n x_n)]$, where $\omega$ is the angular frequency of the wave, $q_n$ denotes the $n$-th component of the wave vector $\vec q$, and the amplitudes $e_j$ and $b_j$ give the electric and magnetic polarization of the wave, respectively. Vectors $e_j$ and $b_j$ can generally be complex quantities. However, in the lossless regime here assumed, they are real. Hereinafter $e_i$ will be denoted as the wave polarization vector. Hence, substituting these plane-wave solutions into Maxwell's equations Eqs.~(\ref{1b}) and (\ref{2b}), together with the constitutive relations Eqs.~(\ref{constitutive1}) and (\ref{constitutive2}), which are taken as time-domain equations (some consequences of including temporal dispersion in the context of the approximations here implemented are briefly discussed in Appendix~\ref{nonlocal}), a straightforward calculation, where $b_j$ is eliminated in favor of $e_j$, leads to the eigenvalue problem, 
\begin{equation}
Z_{ij}e_{j}=0,
\label{eigenvalue}
\end{equation}
where we defined 
\begin{align}
Z_{ij}=&\; C_{ij}v^{2}+\big(\epsilon_{inl}\tilde{H}_{lj}+\epsilon_{lnj}\tilde{C}_{il} \big)\kappa_{n}v 
\nonumber \\
&+ \epsilon_{inl}\epsilon_{kpj}H_{lk}\kappa_{n}\kappa_{p}.
\label{Zij}
\end{align}
In the above result $v$ represents the magnitude of the phase velocity $\vec v =  (\omega/q) \hat\kappa \doteq v \hat \kappa$ \cite{born}, and $\kappa_{i}$ is the $i$-th component of the unit  wave vector $\hat{\kappa}=\vec{q}/q$, so that  $\kappa_i = q_i/q$. Furthermore, we defined the auxiliary optical coefficients, 
\begin{eqnarray}
C_{ij}&=&\varepsilon_{ij}+\frac{\partial\varepsilon_{ik}}{\partial E_{j}}E_{k}+\frac{\partial\tilde{\varepsilon}_{ik}}{\partial E_{j}}B_{k},
\label{C}
\\
\tilde{C}_{ij}&=&\tilde{\varepsilon}_{ij}+\frac{\partial\tilde{\varepsilon}_{ik}}{\partial B_{j}}B_{k}+\frac{\partial\varepsilon_{ik}}{\partial B_{j}}E_{k},
\label{ctilde}
\\
H_{ij}&=&\mu^{\mbox{\tiny (-1)}} _{ij}+\frac{\partial\mu^{\mbox{\tiny (-1)}} _{ik}}{\partial B_{j}}B_{k}+\frac{\partial\tilde{\mu}^{\mbox{\tiny (-1)}} _{ik}}{\partial B_{j}}E_{k},
\label{h}
\\
\tilde{H}_{ij}&=&\tilde{\mu}^{\mbox{\tiny (-1)}} _{ij}+\frac{\partial\tilde{\mu}^{\mbox{\tiny (-1)}} _{ik}}{\partial E_{j}}E_{k}+\frac{\partial\mu^{\mbox{\tiny (-1)}} _{ik}}{\partial E_{j}}B_{k}.
\label{htilde}
\end{eqnarray}
Given the approximations leading to Eq.~(\ref{eigenvalue}), electric and magnetic fields appearing in expressions resulting from Eqs.~(\ref{C}) to (\ref{htilde}) should be approximated by their background contributions $E_i^0$ and $B_i^0$. 

Formally, phase velocities and polarization vectors associated with the wave propagation can be found by solving the eigenvalue problem stated by Eq.~(\ref{eigenvalue}). Suppose the eigenvalues of the matrix whose elements are $Z_{ij}$ are $\lambda_i$ (with $i=1,2,3$), which are functions of $v$. The kernel of this matrix -- the set of eigenvectors corresponding to null eigenvalues -- gives the wave polarization vectors $e_i$. The corresponding phase velocities are obtained by solving the set of equations $\{\lambda_i(v) = 0\}$  for $v$, or equivalently, by solving ${\rm det}(Z_{ij})=0$ for $v$. 

It should be mentioned that the general expression for $Z_{ij}$ in Eq.~(\ref{Zij}) can also be obtained by using Hadamard's method of discontinuities of fields derivatives across a moving hypersurface, as shown in earlier studies  \cite{2012PhRvA..86a3801D,2017PhRvA..95c3826B,*2021PhRvA.104d3523B}.
Furthermore, different approaches can be implemented, such as, for instance, the use of four dimensional notation, which provides a natural scenario to study analogies between general relativity solutions and wave propagation in optical media \cite{2002PhRvE..65b6612D,*2002PhRvD..65f4027D,*2006PhLA..360...10D} or, more generally, in condensed-matter systems --  the so called analog gravity \cite{2005LRR.....8...12B}. The gravitational analog of the linear magnetoelectric effect was recently examined \cite{2019Univ....5...88G}.

\section{Wave propagation in second-order magnetoelectric optical material} 
\label{second-order}
The subject of this section is the study of the influence of the second-order magnetoelectric effect on light propagation in a nonlinear optical medium, which is characterized by the optical coefficients $\beta_{ijk}$. Thus, assuming the absence of spontaneous polarization and magnetization, and keeping only $\chi_{ij}^{\mbox{\tiny $(1)$}}$ and $\tilde\chi_{ij}^{\mbox{\tiny $(1)$}}$ linear electromagnetic contributions, together with the second-order magnetoelectric contribution, the prescription in Appendix~\ref{Polarization} leads to the following polarization and magnetization vectors describing the optical medium, 
\begin{eqnarray}
P_{i}&=&\varepsilon_{0}\chi_{ij}^{\mbox{\tiny $(1)$}}E_{j}+\frac{1}{2}\beta_{ijk} H_{j}H_{k},
\label{Pi2}
\\
M_{i}&=&\tilde\chi_{ij}^{\mbox{\tiny $(1)$}}H_{j}+\frac{1}{\mu_{0}}\beta_{jki}E_{j}H_{k}.
\label{Mi2}
\end{eqnarray}
The assumption of keeping only second-order magnetoelectric terms related to $\beta_{ijk}$ allows us to focus on the effect of these contributions to the propagation of light rays in such media. Notwithstanding, it should be mentioned that materials with such behavior do exist in nature, such as, for instance, the perovskite (${\rm BiFe O}_3$) \cite{1985Tabares} where the linear magnetoelectric contribution is canceled due to its incommensurate structure \cite{doi:10.1080/00150199408213372}. 
Possible effects related to $\gamma_{ijk}$, which also corresponds to a second-order contribution, are not being considered here. Formally, our results are restricted to materials for which such contribution is not permitted \cite{1973-schmid} in the expansion of the free-energy density, presented in Appendix~\ref{Polarization}. Systems for which such contribution is allowed and induce effects of the same magnitude of those related to $\beta_{ijk}$, require further analysis. However, it is important to notice that the $\gamma_{ijk}$ term appearing in the expansion of the free-energy density is coupled to the square of the applied electric field. Therefore, its contribution to polarization and magnetization phenomena will always  depend on the magnitude of this field. Thus, even when this term is allowed, in the absence of an  applied electric field effects related to the $\gamma$-coupling are naturally suppressed. On the other hand, even in the absence of the electric field, effects related to $\beta$-coupling, as stated by Eqs. (\ref{Pi2}) and (\ref{Mi2}),  will still be present.


Given the above expression for the magnetization $M_i$, the magnetic field can be written in terms of the auxiliary field $H_i$ as $B_i = \mu_0(H_i+M_i) =[\mu_0(\delta_{ij}+\tilde\chi_{ij}^{\mbox{\tiny $(1)$}}) + \beta_{kij}E_{k}]H_{j} \doteq A_{ij}H_j$. However it is more convenient to our purposes to express the inverse relation $H_i = A^{\mbox{\tiny -1}}_{ij}B_j$, where $A_{ij}A^{\mbox{\tiny-1}}_{jk} = \delta_{ik}$. Then, keeping only first-order terms in $\beta_{ijk}$ coefficient,   and assuming that the linear magnetic susceptibility is isotropic, i.e., $\tilde\chi_{ij}^{\mbox{\tiny $(1)$}}=\tilde\chi\delta_{ij}$, we obtain that $A^{\mbox{\tiny -1}}_{ij} = (1/\mu) \delta_{ij}-(1/\mu)^2 \beta_{kij}E_{k}$, where we have defined the isotropic magnetic permeability $\mu \doteq \mu_0(1+\tilde\chi)$. Now, returning to the constitutive relations, the following coefficients can be identified, 
\begin{align}
\varepsilon_{ij}& = \varepsilon_0\,(\delta_{ij} + \chi_{ij}^{\mbox{\tiny $(1)$}})\,, 
\label{eij1}
\\
\tilde\varepsilon_{ij} &= \frac{1}{2\mu^{2}}\beta_{ijk}B_k\,,
\label{eij2}
\\
\mu_{ij}^{\mbox{\tiny (-1)}}& = \frac{1}{\mu}\delta_{ij}\,,
\label{mij1}
\\
\tilde{\mu}_{ij}^{\mbox{\tiny (-1)}} & =  -\frac{1}{\mu^2}\beta_{jik}B_k \,.
\label{mij2}
\end{align}
It was recently suggested \cite{2019PhRvA..88841} that an idealized model based on this system could provide a possible scenario for triple-refringence of light in the realm of metamaterials [incidentally, there is a misprint in Eq.~(22) in Ref.~\cite{2019PhRvA..88841} where, instead of $\delta_{ij}$, it should appear $\delta_{iy}\delta_{jy}$]. 

Using the above coefficients in Eqs.~(\ref{C}) to (\ref{htilde}) and inserting the results into Eq.~(\ref{Zij}) it is a straightforward exercise to obtain 
\begin{align}
Z_{ij}=&\varepsilon_{ij}v^2 
+\frac{1}{\mu^2}f_{ij} v - \frac{1}{\mu^2}\big( \mu I_{ij} + a_{ij}\big),
\label{zij}
\end{align}
where $I_{ij} \doteq \delta_{ij} - \kappa_i \kappa_j$ is a two-dimensional projector over the subspace orthogonal to the wave vector $q_i$, so that $I_{ij}I_{jk} = I_{ik}$ and $I_{ij}\kappa_j=0$, and we have defined 
$a_{ij} \doteq ( \delta_{ki}\delta_{lj} - \delta_{kl}I_{ij} + \delta_{ij}\kappa_k\kappa_l - \delta_{li}\kappa_j \kappa_k - \delta_{lj}\kappa_i \kappa_k)   \beta_{nkl}E^0_n$ and  $f_{ij} \doteq (\epsilon_{lki}\beta_{jln}+\epsilon_{lkj}\beta_{iln})B^0_{n}\kappa_k$. 
Furthermore, it is worth noticing that $a_{ij}\kappa_j=0$ and $f_{ij}\kappa_i \kappa_j=0$.

Following the prescription described in Sec.~\ref{wave-propagation}, we now calculate the phase velocities of the light rays that are allowed to propagate in such nonlinear media. The determinant of the $3\times3$ matrix whose elements are $Z_{ij}$ can be obtained by means of its traces $Z_{1}=Z_{ii}$, $Z_{2}=Z_{ij}Z_{ji}$, and $Z_{3}=Z_{ij}Z_{jk}Z_{ki}$, as
$$6\det ( Z_{ij} )=(Z_{1})^{3}-3Z_{1}Z_{2}+2Z_{3}.$$ 

Thus, solving $\det (Z_{ij})=0$ for $v$ leads to the forth degree algebraic equation,
\begin{equation}
\alpha_4 v^{4} + \alpha_3 v^{3} + \alpha_2 v^{2} + \alpha_1 v + \alpha_0 = 0,
\label{quartic}
\end{equation}
where we defined,
\begin{align}
\alpha_4 & =  6\det (\varepsilon_{ij})
\label{a4}
\\
\alpha_3 &= \frac{3}{\mu^2}(\varepsilon_{ik}\varepsilon_{kj}-\varepsilon_{kk}\varepsilon_{ij})(2f_{ij} -f_{ll}\delta_{ij})
\label{a3}
\\
\alpha_2 &= \frac{3}{\mu^2}(\varepsilon_{ik}\varepsilon_{kj}-\varepsilon_{kk}\varepsilon_{ij})(2\mu\kappa_i \kappa_j-2a_{ij}+a_{ll}\delta_{ij}) 
\label{a2}
\\
\alpha_1 &= \frac{6}{\mu^3}(2f_{ij} - f_{ll}\delta_{ij})\varepsilon_{in}\kappa_n \kappa_j
\label{a1}
\\
\alpha_0 &= \frac{6}{\mu^3}(\mu + a_{kk})\varepsilon_{ij}\kappa_i \kappa_j \,.
\label{a0}
\end{align}
In obtaining Eq.~(\ref{quartic}) it is perhaps helpful to notice that $\det (\mu I_{ij} + a_{ij}) = 0$. As one can see, $\alpha_0$ and $\alpha_2$ include contributions that couples the magnetoelectric coefficients $\beta_{ijk}$ with the external electric field, while in $\alpha_1$ and $\alpha_3$ the coupling is with the external magnetic field. Particularly, in the absence of a magnetic field these latter coefficients vanish, as $f_{ij}$ is zero in such case, and Eq.~(\ref{quartic}) reduces to a biquadratic equation for $v$. This is also the case when only linear effects are considered. 

\section{Nonlinear medium presenting an isotropic linear-susceptibility sector}
\label{iso}
\subsection{Phase and group velocities}
Let us investigate the special case where the nonlinear magnetoelectric material has isotropy in its linear susceptibility sector. In such media we have $\chi_{ij}^{\mbox{\tiny $(1)$}} = \chi\, \delta _{ij}$, which leads to $\varepsilon_{ij} = \varepsilon_0(1+\chi)\, \delta_{ij}\doteq \varepsilon\, \delta_{ij}$, where $\varepsilon$ is identified as the isotropic electric permittivity, and the nonlinear sector is still described by Eqs.~(\ref{eij2}) and (\ref{mij2}). Now, $\det (\varepsilon_{ij}) = \varepsilon^3$ and it is a simple task to show that the coefficients in Eq.~(\ref{quartic}) reduce to $\alpha_4  =  6 \varepsilon^3$, $\alpha_3  =  6 (\varepsilon/\mu)^2 f_{ii}$, $\alpha_2  =  -6 (\varepsilon/\mu)^2 ( a_{ii} + 2\mu)$, $\alpha_1  =  -6 (\varepsilon/\mu^3) f_{ii}$, and $\alpha_0  =  6 (\varepsilon/\mu^3)(a_{ii} + \mu)$. Thus, inserting these coefficients in Eq.~(\ref{quartic}), it results in the following solutions,
\begin{align}
&v^{\pm}_{o}=\pm\frac{1}{\sqrt{\varepsilon\mu}},
\label{o-iso}
\\
&v^{\pm}_{e}=-\frac{\epsilon_{iln}\beta_{ilk}B^0_k\kappa_n}{\varepsilon\mu^2}    \pm \frac{1}{\sqrt{\varepsilon\mu}}\left(1 -\frac{1}{2\mu} \beta_{kij}E^0_kI_{ij} \right) .
\label{e-iso}
\end{align}

As it is clear, the solution $v_o^\pm$ corresponds to the phase velocity of an ordinary ray, as its magnitude does not depend on the direction of the wave propagation, i.e., it is isotropic. Additionally, this solution does not depend on the magnetoelectric properties of the medium as well, but only on the linear contribution to the electric and magnetic susceptibilities. On the other hand, the solution $v^{\pm}_{e}$ is associated with the extraordinary ray and depends on the direction of the wave vector and also on the nonlinear contributions related to the magnotoelectric coefficients described by $\beta_{ijk}$. These solutions show that birefringence is activated and controlled by applied external fields, which are coupled to the nonlinear magnetoelectric properties of the material. When both electric and magnetic fields vanish, solutions $v^{\pm}_{e}$ and $v^{\pm}_{o}$ will coincide in any direction of propagation. In this sense, the effect is artificially activated by the external fields in a similar way as occurs in Kerr or Cotton-Mouton effects. 

\begin{figure}[t!]
\center
\includegraphics[width=0.48\textwidth]{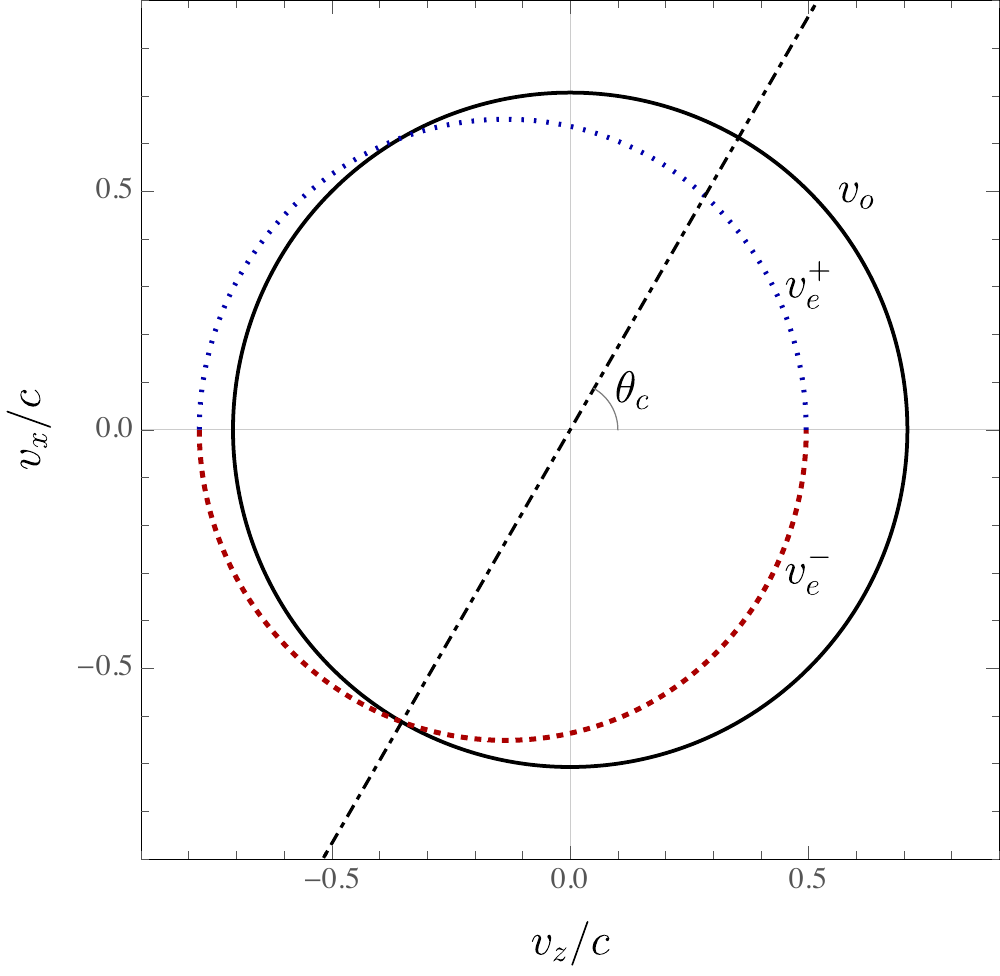}
\caption{Normal surfaces \cite{born} of the magnetoelectric medium characterized by  Eqs.~(\ref{eij1}) to (\ref{mij2}), with $\chi_{ij}^{\mbox{\tiny $(1)$}} = \chi\, \delta_{ij}$. It is assumed that $\beta_{122}$ is the only  non-negligible component of $\beta_{ijk}$, which couples to both applied electric and magnetic fields.  Due to the presence of the magnetic field, birefringence is not symmetric under space reversal.}
\label{fig122}
\end{figure}

\begin{figure}[t!]
\center
\includegraphics[width=0.48\textwidth]{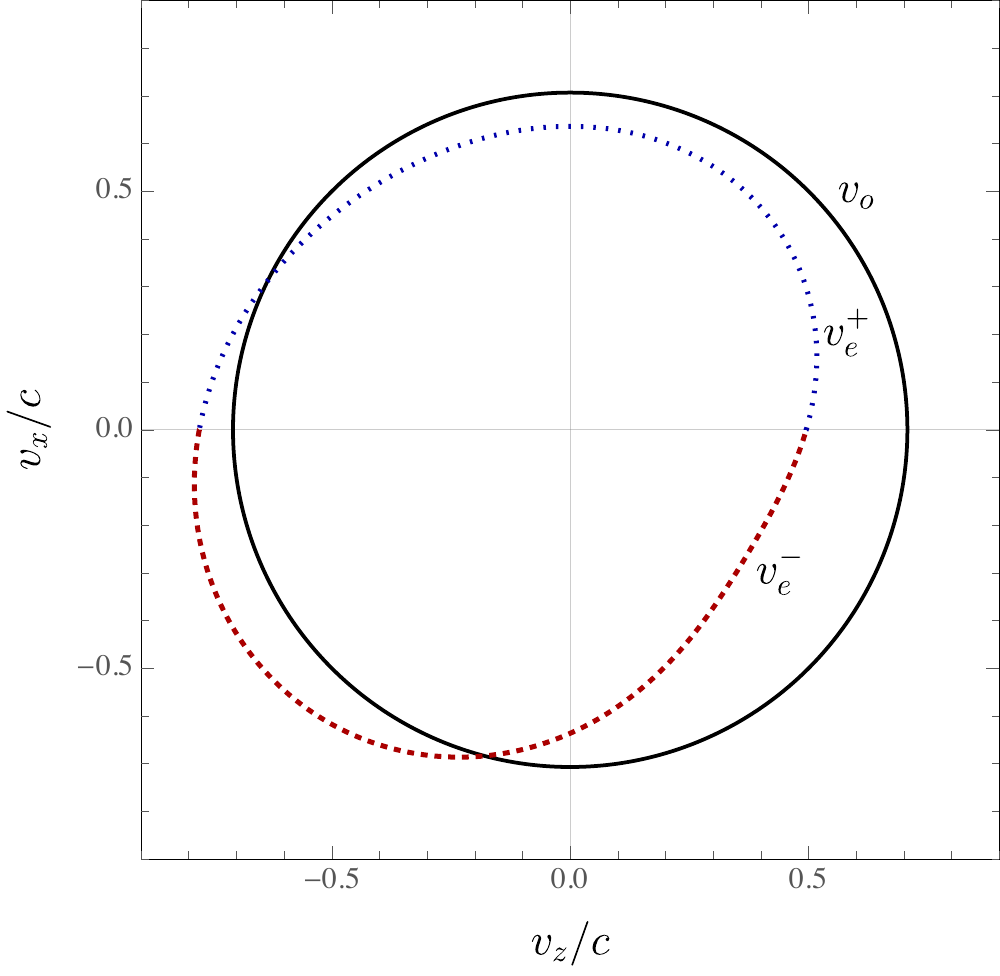}
\caption{Normal surfaces \cite{born} of the magnetoelectric medium characterized by  Eqs.~(\ref{eij1}) to (\ref{mij2}), with $\chi_{ij}^{\mbox{\tiny $(1)$}} = \chi\, \delta_{ij}$. Here it is assumed that $\beta_{113}$ and $\beta_{122}$ are the only  non-negligible components of $\beta_{ijk}$. Both,  applied electric and magnetic fields are present. Compared with Fig.~\ref{fig122} the effect of including $\beta_{113}$ is to produce a distortion of the extraordinary solution. }
\label{fig113122}
\end{figure}

\begin{figure}[t!]
\center
\includegraphics[width=0.48\textwidth]{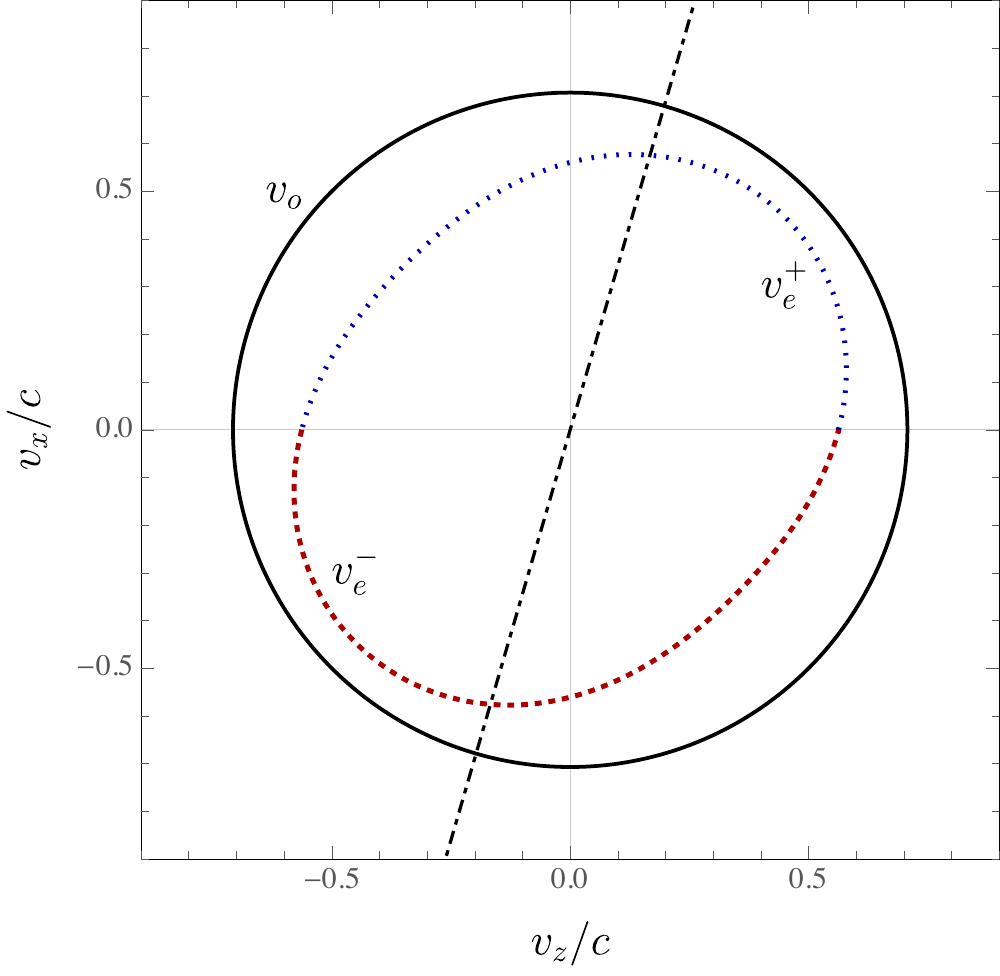}
\caption{Normal surfaces \cite{born} of the magnetoelectric medium characterized by  Eqs.~(\ref{eij1}) to (\ref{mij2}), with $\chi_{ij}^{\mbox{\tiny $(1)$}} = \chi\, \delta_{ij}$. Here it is assumed that all components of  $\beta_{ijk}$ contribute to the effect, but only an applied electric field is present. In this case, due to the absence of a magnetic field, the birefringence effect exhibits symmetry under space reversal (see, for instance, the distance between ordinary and extraordinary curves along the straight dot-dashed line).}
\label{figall}
\end{figure}

To numerically examine some illustrative models, suppose for simplicity that electric and magnetic fields are applied in such way that $E^0_i = E \delta_{i1}$, $B^0_i = B \delta_{i2}$, where $E$ and $B$ are the magnitude of these constant fields, and the propagation direction is set in an arbitrary direction in the $XZ$ plane, i.e., $\kappa_i(\theta) = \delta_{i1} \sin\theta  + \delta_{i3} \cos\theta$, with $\theta$ measured from the $v_z/c$ axis. In such case, the phase velocities of the ordinary and the extraordinary rays can be calculated as functions of the optical coefficients $\varepsilon$, $\mu$,  and $\beta_{ijk}$. 
Possible arrangements presenting birefringence are depicted in Figs.~\ref{fig122}, \ref{fig113122} and \ref{figall}, where we set $v_o =c/\sqrt{2}$,  $\beta E/(2\mu) \approx 0.1$, and $v_o \beta B/\mu\approx 0.2$, with $\beta$ denoting the considered coefficient $\beta_{ijk}$ mentioned in the caption of each figure. In these figures, distances from the origin to the curves give the magnitude of the corresponding phase velocities in the wave-vector direction $\hat \kappa(\theta)$. 

In Fig.~\ref{fig122} it is assumed that $\beta_{122}$ is the only one non-negligible component of $\beta_{ijk}$. Both applied fields are coupled to this coefficient and contribute to the phenomenon. Notice that the effect is nonsymmetric under space reversal, which means that the velocity of the extraordinary ray will be distinct in $\hat \kappa(\theta)$ and $\hat\kappa(\theta+\pi) = -\hat \kappa(\theta)$ directions. In particular, we see that there are some specific directions for which both ordinary and extraordinary rays have the same phase velocity, such as, for instance, occurs with $v_e^-$ and $v_o$ in the direction indicated by the dot-dashed line in this figure. Furthermore, in this specific direction birefringence occurs only in one side, corresponding to the positive roots of the phase velocities, i.e., there will be birefringence in  $\hat\kappa(\theta_c)$ direction but only single-refringence in $\hat\kappa(\theta_c+\pi)$ direction. 
The polarization vectors associated with  ordinary and extraordinary rays can be calculated by means of the method described in Sec.~\ref{wave-propagation}. For instance, assuming the simplified model investigated in Fig.~\ref{fig122}, the wave polarization vector of the ordinary ray does not depend on direction of propagation in the $XZ$-plane, and its direction is given by $(0,1,0)$.  On the other hand, the wave polarization vector of the extraordinary ray depends on the direction of the wave propagation $\hat\kappa$. It can be calculated for an arbitrary direction, but the easiest way to obtain this vector is by choosing each specific direction of propagation of interest. For instance, if we set propagation in the $X$-direction, which corresponds to choosing $\kappa_i = \delta_{i1}$, i.e., setting $\theta=\pi/2$,  the corresponding normalized wave polarization vector will be given by $(0,0,1)$. Thus, we have the classical picture of the birefringence phenomenon, where two light rays (ordinary and  extraordinary ones) propagate in the same direction with different phase velocities, each ray presenting a different (and linearly independent) wave polarization vector. Similarly, if the propagation is set in the $Z$-direction ($\kappa_i = \delta_{i3}$) it can be shown that the corresponding  normalized wave polarization vector of the extraordinary ray will be given by $(1,0,0)$.

The effect of assuming other significant components of $\beta_{ijk}$ is just to distort the shape of the normal surfaces related to the extraordinary ray, as shown in Fig.~\ref{fig113122} (compare with Fig.~\ref{fig122}), where a model is depicted for which both $\beta_{122}$ and $\beta_{113}$ are significant, with all other components  assumed to be negligible. 

There is one aspect that is worth mentioning. If only the electric field is present, no matter which components of $\beta_{ijk}$ are considered, positive and negative roots of $v^{\pm}_{e}$ will have the same magnitude. In this case $|v^{\pm}_{e}(\hat \kappa)| = |v^{\pm}_{e}(-\hat \kappa)|$ and birefringence is symmetric under space reversal, as it can be seen in the model depicted in Fig.~\ref{figall}. 

We close this section by presenting the expressions for the group velocities associated with wave propagation in this class of magnetoelectric material. The group velocity can be obtained by means of,  
\begin{align}
&\vec u = \frac{\partial \omega}{\partial \vec q}  =v \hat \kappa + q\frac{\partial v}{\partial \vec q}\;\,.
\label{vg}
\end{align}
Using the positive roots of the phase velocities given by Eqs.~(\ref{o-iso}) and (\ref{e-iso}), it is not difficult to show in the present case that the phase and group velocities have the same magnitudes, i.e., $v = u$, for ordinary and extraordinary solutions. However, they generally differ in direction. While the ordinary wave is such that $\vec u_o = v_o \hat \kappa = \vec v_o$, the group velocity associated with the extraordinary wave is given by $\vec u_e = v_e \hat u_e$, where the unit vector $\hat u_e$ is given by, 
\begin{align}
\hat u_e = \hat \kappa + \frac{1}{\mu} \left( \beta_{kij}E^0_k \kappa_i+ v_o \epsilon_{lij}\beta_{ilk}B^0_k \right)I_{jn}\hat x_n\,,
\label{vge}
\end{align}
with $\hat x _n$ denoting the unit vector in the $n$-th direction in Cartesian coordinates, so that $\vec q \cdot \hat x_n = q_n$. Furthermore, we used the relation $q (\partial \kappa_i /\partial \vec q\,) = \hat x_i - \kappa_i \hat\kappa = I_{ij} \hat x_j$. Thus, due to the nonlinear properties of the medium the direction of the group velocity associated with the extraordinary wave will not coincide with the direction given by the wave vector.

\subsection{Obtaining the $\beta_{ijk}$ coefficients by means of birefringence measurements}
\label{beta}
Let us now examine a method of obtaining the magnetoelectric coefficients by means of measurement of the refractive indexes of the medium. The possibility of measuring second-order magnetoelectric coefficients by means of birefringence effects were already reported \cite{1987PhR...155..379E} in the context of $\gamma_{ijk}$ contribution.

The refractive index is generically defined by $n\doteq c/v$. If we choose the positive roots of the phase velocities in Eqs.~(\ref{o-iso}) and (\ref{e-iso}), we obtain the refractive indexes experienced by the ordinary $n_o$ and extraordinary $n_e$ rays, respectively,
\begin{align}
&n_{o}=c \sqrt{\varepsilon\mu},
\label{no-iso}
\\
&n_{e}=n_o\left( 1 + \frac{1}{2\mu} \beta_{kij}E^0_kI_{ij}  + \frac{v_o}{\mu} \epsilon_{lni}\beta_{ilk}B^0_k\kappa_n   \right) .
\label{ne-iso}
\end{align}
Notice that these quantities are dependent on the direction of propagation set by the wave vector. Now, we define the fractional deviation of the refractive index as $\Delta \doteq (n_e-n_o)/n_o$, and we denote by $\Delta_i$ the value of this deviation when the propagation is set in the $i$-th direction. For instance, $\Delta_x = \Delta(\kappa_i = \delta_{i1})$. Hence, calculating $\Delta_i$ for the three orthogonal directions we obtain,
\begin{align}
&\Delta_x  = \frac{1}{2\mu}\left[\left(\beta_{j22}+\beta_{j33}\right) E^0_j + \frac{2c}{n_o}\left(\beta_{23j}-\beta_{32j}\right) B^0_j\right]
\label{dx}
\\
&\Delta_y  = \frac{1}{2\mu}\left[\left(\beta_{j11}+\beta_{j33}\right) E^0_j + \frac{2c}{n_o}\left(\beta_{31j}-\beta_{13j}\right) B^0_j\right]
\label{dy}
\\
&\Delta_z  = \frac{1}{2\mu}\left[\left(\beta_{j11}+\beta_{j22}\right) E^0_j + \frac{2c}{n_o}\left(\beta_{12j}-\beta_{21j}\right) B^0_j\right]
\label{dz}
\end{align}
Thus, conveniently setting the directions of the applied electric and magnetic fields and measuring the deviations $\Delta_i$ (for the three directions $i =x,y,z$), the magnetoelectric coefficients can be obtained directly from the above system of equations, as shown bellow. 

Suppose an experiment is carried out in such a way that the electric field is set in the $X$-direction such that $E^0_i = E \delta_{i1}$, and the magnetic field is turned off ($B^0_i = 0$). Thus, after measuring the deviations $\Delta_i$, Eqs. (\ref{dx}) to (\ref{dz}) can be solved to obtain the following components of the beta coefficients of the material,
\begin{align}
&\beta_{111} = \frac{\mu}{E}\left(-\Delta_x+\Delta_y+\Delta_z\right) ,
\nonumber
\\
&\beta_{122} = \frac{\mu}{E}\left(\Delta_x-\Delta_y+\Delta_z\right) ,
\nonumber
\\
&\beta_{133} = \frac{\mu}{E}\left(\Delta_x+\Delta_y-\Delta_z\right) .
\nonumber
\end{align}
More generally, if the deviations are measured with the electric field set in the $n$-th direction, $n=1,2,3$, the coefficients $\beta_{n11}$, $\beta_{n22}$ and $\beta_{n33}$, will be obtained in the same way. 

Now, once the nine components above of $\beta_{ijk}$ are known, almost all the other components can be obtained by setting $E^0_i =0$ and playing with the directions of the wave vector $\vec q$ and the applied magnetic field in Eqs.~(\ref{dx}) to (\ref{dz}). 
Setting $E^0_j =0$ and $B^0_j=B \delta_{j1}$ in Eqs.~(\ref{dy}) and (\ref{dz}) leads to
\begin{align}
\beta_{112} &=  \beta_{211} + \frac{\mu n_o}{cB} \Delta_z,
\nonumber\\
\beta_{113} &=  \beta_{311}  - \frac{\mu n_o}{cB} \Delta_y.
\nonumber
\end{align}
On the other hand, setting $E^0_j =0$ and $B^0_j=B \delta_{j2}$ in Eqs.~(\ref{dx}) and (\ref{dz}) leads to
\begin{align}
\beta_{221} &=  \beta_{122} - \frac{\mu n_o}{cB} \Delta_z,
\nonumber\\
\beta_{223} &=  \beta_{322}  + \frac{\mu n_o}{cB} \Delta_x.
\nonumber
\end{align}
Or yet, setting $E^0_j =0$ and $B^0_j=B \delta_{j3}$ in Eqs.~(\ref{dx}) and (\ref{dy}) result in
\begin{align}
\beta_{331} &=  \beta_{133}  + \frac{\mu n_o}{cB} \Delta_y,
\nonumber\\
\beta_{332} &=  \beta_{233}  -  \frac{\mu n_o}{cB} \Delta_x.
\nonumber
\end{align}

The components $\beta_{123}$, $\beta_{213}$ and $\beta_{312}$ are the only ones that cannot be obtained independently in terms of deviations measurements, but in terms of their differences,
\begin{align}
\beta_{213}  -  \beta_{312}  &=  \frac{\mu n_o}{cB} \Delta_x, \;\;\;\; ( E^0_i =0,  B^0_i=B\delta_{i1})
\nonumber\\
\beta_{312}  -  \beta_{123}  &=  \frac{\mu n_o}{cB} \Delta_y, \;\;\;\; ( E^0_i =0,  B^0_i=B\delta_{i2})
\nonumber\\
\beta_{123}  -  \beta_{213}  &=  \frac{\mu n_o}{cB} \Delta_z, \;\;\;\; ( E^0_i =0,  B^0_i=B\delta_{i3}).
\nonumber
\end{align}

As it can be concluded, birefringence seems to be an interesting tool to measure the components of the second-order magnetoelectric coefficients $\beta_{ijk}$ for this class of optical material.
%
%
%

\subsection{Estimates for birefringence measurements}
\label{estimates}
Measurements of second-order magnetoelectric effects have been reported on several crystals and the related coefficients were estimated. Charge integration technique is normally used to obtain the coefficients and, most commonly, the values obtained lie in the range $(10^{-17} -10^{-19})$  ${\rm s\,A^{-1}}$. Just to mention some results, the effect was reported in ${\rm NiSO}_4\cdot6\rm{H}_2\rm{O}$ ($\beta_{223} = -2.75\times10^{-17} {\rm s\,A}^{-1}$ at 4.2K)~\cite{1965PhRv..138.1218H},  ${\rm BiFeO}_3$ ($\beta_{113} = 8.1\times10^{-19} {\rm s\,A}^{-1}$ at 4K) ~\cite{1985Tabares}, and ${\rm Cr}_3{\rm B}_7\rm{O}_{13}{\rm Cl}$ ($\beta_{333} = 1.5\times10^{-18} {\rm s\,A}^{-1}$ at 4.2K)~\cite{doi:10.1080/00150199408213358}.

As discussed in Sec.~\ref{beta}, such coefficients could be obtained by measurements of birefringence effect. Selecting the X-direction for the wave-vector, an estimate of the effect can be obtained by means of,
\begin{align}
\Delta_x = \frac{\mu_0}{\mu}&\left[3.98\times 10^{-7}\left(\frac{\beta_{j22}+\beta_{j33}}{10^{-17}{\rm s\, A}^{-1}}\right)\left(\frac{E^0_j}{10^5 {\rm V\,m}^{-1}}\right) \right.
\nonumber \\
&\left. + 1.19\times 10^{-3}\left(\frac{2}{n_o}\right)\left(\frac{\beta_{23j}-\beta_{32j}}{10^{-17}{\rm s \,A}^{-1}}\right)\left(\frac{B^0_j}{1{\rm T}}\right) \right].
\nonumber
\end{align}
Similar expressions hold for $Y$ and $Z$ directions. It is interesting to notice that  the contribution coupled to the magnetic field is stronger than the one coupled to the electric field, even when we set a large reference value for the latter, as it is clear in the above expression. 

To deal with a specific case, if we set up an experiment where the magnetic field is applied in the $Y$ direction, with $B=1 {\rm T}$, for a system with $\mu \approx \mu_0$, $n_o \approx 2$, and in the absence of an external electric field, we get
\begin{align}
\Delta_x (E^0_j=0; B^0_j=B\delta_{j2})\approx 1.19\times 10^{-3} \left(\frac{\beta_{232}-\beta_{322}}{10^{-17}{\rm s \,A}^{-1}}\right).
\nonumber
\end{align}
If the coefficients $\beta_{ijk}$ are of the order of $10^{-17}{\rm s\,A}^{-1}$, we conclude that a deviation of the refractive index of the order of $10^{-3}$ would be expected. This seems to be an effect easily detectable in modern optics laboratories. Even effects coupled to the electric field ($\approx 10^{-7}$) are certainly within the current measurability range. 

Nonreciprocity of light can be estimated in a similar way by comparing $\Delta_i(\hat \kappa)$ with $\Delta_i(-\hat \kappa)$. It is interesting to note that there may be situations in which birefringence occurs in only one direction, as unveiled in the idealized configurations depicted in Figs. \ref{fig122} and \ref{fig113122}.

Alternatively, in terms of measurements of phase velocity, a possible observable for this effect could be obtained by means of the fractional difference,
\begin{align}
\frac{| v_e(\hat \kappa) - v_e(-\hat\kappa)|}{c} \approx\,& 4.77\times 10^{-3}\left(\frac{\varepsilon_0}{\varepsilon}\right) \left(\frac{\mu_0}{\mu}\right)^2 
\nonumber \\ 
&\times \epsilon_{ijl}\kappa_l \left(\frac{\beta_{ijn}}{10^{-17}{\rm s\, A}^{-1}}\right)\left(\frac{B_n^0}{1 {\rm T}}\right),
\nonumber
\end{align}
where $c$ is the speed of light in vacuum. In this expression, once the directions of the wave vector and the applied magnetic field are adjusted, an estimate of the effect can be obtained. Notice that only the magnetic field appears in this expression. As can be seen, for typical values of the above parameters, a fractional difference of the phase velocity of about $10^{-3}$ is expected.

\section{Some aspects about wave propagation in materials presenting a natural optic axis}
\label{aniso}
Suppose now that the magnetoelectric medium described by Eqs.~(\ref{eij1}) to (\ref{mij2}) possesses a natural optic axis in the $X$-direction, so that $\varepsilon_{ij} = {\rm diag}  (\varepsilon_{\parallel},\varepsilon_{\perp},\varepsilon_{\perp})$. In this case, up to a global factor $6\varepsilon_\perp^2/\mu^2$, the coefficients in Eqs.~(\ref{a4}) to (\ref{a0}) reduce to,
\begin{align}
\alpha_4 &= \mu^2\varepsilon_\parallel\,,
\nonumber
\\
\alpha_3 &=(1-\gamma)f_{11} +\gamma f_{ii} \,,
\nonumber
\\
\alpha_2 &= \mu \left[(1-\gamma)\kappa_1^2-1 - \gamma\right] - (1-\gamma)a_{11} -\gamma a_{ii} \,,
\nonumber
\\
\alpha_1 &= \frac{1}{\mu\varepsilon_\perp}\left\{\left[(1-\gamma)\kappa_1^2-1\right] f_{ii} + \frac{2}{\varepsilon_\perp} \varepsilon_{ij}f_{jn}\kappa_n \kappa_i\right\} \,,
\nonumber
\\
\alpha_0 &= -  \frac{1}{\mu\varepsilon_\perp}\left[(1-\gamma)\kappa_1^2-1\right](a_{ii}+\mu) \,,
\nonumber
\end{align}
where we have defined $\gamma \doteq \varepsilon_\parallel/\varepsilon_\perp$.
Solutions for the phase velocity in such medium can be found by introducing the above coefficients into Eq.~(\ref{quartic}). However, it is not our interest here to study the general case, but only to address some representative particular configurations. 

First, it is worth recapping that in the absence of applied electric and magnetic fields the nonlinear contributions are ``turned-off", and we obtain the well-known case of linear media presenting an optic axis. In this case, setting $a_{ij}$ and $f_{ij}$ to zero and solving Eq.~(\ref{quartic}) to the phase velocity $v$, we obtain that there will be an ordinary ray with $v_o = \pm (\mu\varepsilon_\perp)^{-1/2}$ and an extraordinary ray with $v_{e}^{\pm} = \pm (\mu\varepsilon_\parallel)^{-1/2}[1+(\varepsilon_\parallel/\varepsilon_\perp -1)\kappa_1^2]^{1/2}$. Notice that there will be no birefringence effect when the propagation is set in the direction given by the optic axis, i.e., setting $\kappa_i =\delta_{i1}$, as expected, and the effect occurs with maximum magnitude when the propagation is set perpendicularly to the optic axis, i.e., when $\kappa_1 = 0$. 

Second, as already mentioned at the end of Sec.~\ref{second-order}, when only an applied electric field is present the fourth-degree equation for the phase velocity reduces to a bi-quadratic equation, $\alpha_4 v^{4} + \alpha_2 v^{2} + \alpha_0 = 0$, 
whose solutions can be used to obtain the components of $\beta_{ijk}$ in terms of measurements of birefringence, similarly to what we have done in the last section.

Finally, when the analysis is extended to the case when an applied magnetic field is present, obtaining simple analytical solutions for the phase velocities can be intricate. Just to explore a convenient model, let us suppose the absence of an electric field, and set the magnetic field perpendicularly to the direction of the natural optic axis, $B^0_i = B \delta_{i2}$. In this case, as $E^0_{i} =0$, all the components of $a_{ij}$ will vanish. Additionally, if the propagation is set along the natural optic axis, i.e., $\kappa_i = \delta_{i1}$, it is straightforward to show that the phase velocities of the ordinary and extraordinary rays in such naturally anisotropic media will be given by,
\begin{align}
&v^{\pm}_{o}=\pm\frac{1}{\sqrt{\varepsilon_\perp \mu}}\, ,
\label{o-anisoX}
\\
&v^{\pm}_{e}= \pm \frac{1}{\sqrt{\varepsilon_\perp \mu}} + \frac{(\beta_{322}-\beta_{232})B}{\varepsilon_\perp \mu^2}    \, .
\label{e-anisoX}
\end{align}
Hence, when the magnetoelectric class of materials here considered is under the influence of an applied magnetic field, birefringence effects can be found even in the direction of the natural optic axis. It is worth noticing that in this direction the phase velocities of both rays do not depend on $\varepsilon_\parallel$. 
The refractive indexes experienced by these rays are different, as they have distinct phase velocities in the same direction. In this particular configuration, the difference between them in such a direction is given by $n_e-n_o = (cB/\mu)\big(\beta_{322} -\beta_{232}\big)$. Thus, as anticipated, the measurement of this deviation provides an indirect measurement of some components of  $\beta_{ijk}$. 
The same analysis can be applied to other directions as well. A more detailed treatment of this system, as well as other similar systems, deserves further investigation.

\section{Final remarks}
\label{final}
Summarizing, some aspects of light propagation in second-order nonlinear magnetoelectric materials were investigated. The analysis was restricted to systems presenting a linear dielectric sector whose polarization and magnetization are coupled to electric and magnetic fields, respectively, and a second-order magnetoelectric sector. It was shown that the magnetoelectric sector contributes in a fundamental way to the occurrence of birefringence phenomena. In particular, birefringence in systems presenting an isotropic linear sector is only activated by external fields by means of the magnetoelectric couplings. On the other hand,  in the case of systems presenting a natural optic axis, the nonlinear couplings are responsible for the induction of birefringence even in the direction of the optic axis, so it breaks a natural symmetry of the system. Furthermore, it was shown that measurements of  birefringence effects can be used as a possible tool for obtaining the components of the second-order magnetoelectric coefficient $\beta_{ijk}$ of the medium. Group velocity was also briefly discussed and it was shown that, in the case of systems presenting an isotropic linear-dielectric-sector, phase and group velocities coincide in magnitude, but not in the direction of propagation.  However, in the absence of external fields the nonlinear couplings are turned-off and the medium behaves just as a linear one with coinciding phase and group velocities. 

It is well known that time-reversal and space-inversion symmetries can be broken in certain magnetoelectric materials. This means that the optical properties of the medium are not the same ones as experienced by a light ray propagating in opposite directions. This effect can be mathematically understood by direct inspection of the phase velocity solutions for the extraordinary ray presented in Secs. \ref{iso} and \ref{aniso}. Implementing $\hat \kappa \to -\hat \kappa$ changes the magnitude of the phase velocities, which means that the light rays experience different refractive indexes in opposite directions. This is a possible scenario for measuring nonreciprocity refraction of light \cite{PhysRevLett.123.077401}. Notice that such behavior is intrinsically related to the coupling between the externally applied magnetic field and the nonlinearities of the medium. When $B_i=0$ the solutions become symmetric under space-inversion. Furthermore, in the absence of this field the algebraic equation that leads to the phase velocities is biquadratic and thus produces only symmetric solutions, thus preserving space-inversion symmetry. 

When the linear contribution to the magnetoelectric effect is present ($\alpha_{ij} \neq 0$), the above-mentioned symmetries are naturally broken. 
However, when the system does not present the linear magnetoelectric effect, and as the nonlinear terms always appear coupled to the applied fields, the presence of one or both of these fields determines if a symmetry is broken. In particular, with regard to phase velocities, the magnetic field is the only field responsible for breaking the space inversion symmetry.

Looking at the phase velocity solutions in Eqs. (\ref{o-iso}) and (\ref{e-iso}), when the magnetic field is present, no matter whether an electric field is also present or not, it seems that there is a possibility of finding three different solutions in one direction, and only one in the opposite direction. However, this is not a real possibility in the scenario examined in this work because the nonlinear contribution coupled to $\beta_{ijk}$ was assumed to be smaller than the linear contribution in the free-energy expansion. This means that $(\epsilon\mu)^{-1/2}$ is the dominant contribution in this solution. Furthermore, only first-order terms in $\beta_{ijk}$ coefficients were kept in the calculations leading to such solutions.  Hence, there will be at most two positive and two negative independent solutions, which means that only single-refringence or  birefringence can occur, as discussed in this work. In spite of this fact, this mathematical aspect suggests that an exact model in which three independent light rays propagate in a same direction with distinct phase velocities could be imagined, and possibly artificially tailored. This observation motivates a possible scenario where a triple-refringence could be artificially produced in the realm of metamaterials. Some idealized models exploring this idea were recently discussed \cite{2012PhRvA..86a3801D,2019PhRvA..88841}. This is a subject that still deserves further investigation. 


\begin{acknowledgments}
Alice V. De Lorenci is acknowledged for reading the manuscript.
This work was partially supported by the Brazilian research agency CNPq (Conselho Nacional de Desenvolvimento Cient\'{\i}fico e Tecnol\'ogico) under Grant No. 305272/2019-5.
\end{acknowledgments}

\appendix
\section{Polarization and magnetization vectors in a nonlinear optical material in a state of thermodynamic equilibrium}
\label{Polarization}
A material medium has its optical properties characterized by certain coefficients that describe how its polarization $P_i$ and magnetization $M_i$ behave when externally applied fields are present. An interesting way of obtaining these quantities is by means of the free-energy density $F(\vec E, \vec H;T)$ of the medium ($T$ holds for temperature), which can be expanded in terms of the electromagnetic fields as \cite{1973-schmid,rivera1994a,2009EPJB...71..299R,2005JPhD...38R.123F},
\begin{eqnarray}
F(\vec{E},\vec{H};T)=&F_{0}& - P^{S}_{i}E_{i} - \mu_0 M^{S}_{i}H_{i}
\nonumber
\\
&-& \frac{1}{2}\varepsilon_{0}\chi_{ij}^{\mbox{\tiny $(1)$}}E_{i}E_{j} -\frac{1}{2}\mu_{0}\tilde\chi_{ij}^{\mbox{\tiny $(1)$}}H_{i}H_{j} - \alpha_{ij}E_{i}H_{j}
\nonumber
\\
&-& \frac{1}{2}\beta_{ijk}E_{i}H_{j}H_{k}-\frac{1}{2}\gamma_{ijk}H_{i}E_{j}E_{k}
\nonumber
\\
&-& \frac{1}{3}\varepsilon_{0}\chi^{{}_{(2)}}_{ijk}E_{i}E_{j}E_{k}-\frac{1}{3} \mu_{0}\tilde\chi^{{}_{(2)}}_{ijk}H_{i}H_{j}H_{k} + ...
\nonumber
\end{eqnarray}
where $F_{0}$ is the free-energy density in the absence of external excitations. 
Before talking about the other coefficients appearing in this expansion, let us derive the polarization $P_i$ [C\,m$^{-2}$] and magnetization $M_i$ [A\,m$^{-1}$] vectors of the material medium, which can be obtained by means of the derivatives of $F$ with respect to $E_i$ and $H_i$ fields as follows,
\begin{align}
P_{i}=&-\frac{\partial F}{\partial E_{i}} 
\nonumber\\
=& \,P_i^S + \varepsilon_{0}\left\{\chi_{ij}^{\mbox{\tiny $(1)$}}E_{j} +\chi^{{}_{(2)}}_{ijk}E_{j}E_{k} + ...\right\} 
\nonumber \\
&+ \left\{\alpha_{ij}H_{j}+\frac{1}{2}\beta_{ijk}H_{j}H_{k}+\gamma_{ijk}H_{k}E_{j} + ...\right\}
\nonumber
\end{align}
and 
\begin{align}
\mu_0 M_{i}=&-\frac{\partial F}{\partial H_{i}}
\nonumber\\
=&\, \mu_0 M_i^S + \mu_0 \left\{\tilde\chi_{ij}^{\mbox{\tiny $(1)$}}H_{j} + \tilde\chi^{{}_{(2)}}_{ijk}H_{j}H_{k} + ...\right\} 
\nonumber \\
&+ \left\{\alpha_{ji}E_{j}+\beta_{kij}E_{k}H_{j}+\frac{1}{2}\gamma_{ijk}E_{j}E_{k} + ...\right\},
\nonumber
\end{align}
where curly brackets were introduced to identify the contributions related to dielectric (the first brackets in $P_i$), magnetic (the first bracket in $M_i$), and magnetoelectric (second brackets in both expressions) effects.
In the above results $P^{S}_{i}$ and $M^{S}_{i}$ are the $i$-th components of the spontaneous polarization and magnetization, respectively, and they are related to zero-th order effects. 
The order of the effects refers to the power of the fields that appear in the above polarization and magnetization vectors.  
Dimensionless coefficients  $\chi^{\mbox{\tiny $(1)$}}_{ij}$  and $\tilde\chi^{\mbox{\tiny $(1)$}}_{ij}$  are the first-order induced electric and magnetic linear contributions to the susceptibilities of the medium, respectively, while $\alpha_{ij}$ [s\,m$^{-1}$] are linear magnotoelectric coefficients --- they are related to a polarization effect linearly induced by a magnetic field or a magnetization effect linearly induced by an electric field. Such effects are called linear magnetoelectric effects. 
%
All the other terms are related to nonlinear processes described by the upper order electric, magnetic and magnetoelectric coefficients in the above expansion. For instance, a medium with non null  $\beta_{ijk}$ [s\,A$^{-1}$] coefficients will exhibit a contribution to the effective dielectric permittivity that depends on the magnitude and direction of the externally applied magnetic field and, at same time, a contribution to the effective magnetic permeability that depends on the magnitude and direction of the externally applied electric field, and  so on.  A similar reasoning applies for $\gamma_{kij}$ [s\,V$^{-1}$].
The last two terms presented in the free-energy density expansion are related to electro-optic and magneto-optic effects, respectively. The corresponding coefficients are the second-order electric $\chi^{{}_{(2)}}_{ijk}$ [m\,V$^{-1}$] and magnetic  $\tilde\chi^{{}_{(2)}}_{ijk}$ [m\,A$^{-1}$] susceptibilities.
 All these coefficients can depend on temperature. 

In the static regime, due to the contraction of $\chi_{ij}^{\mbox{\tiny $(1)$}}$, $\tilde \chi^{\mbox{\tiny $(1)$}}_{ij}$, $\beta_{kij}$ and $\gamma_{kij}$ with squared electric or magnetic fields, only the symmetric part (in the last two indices) of these quantities will survive, as explicitly shown in the expression for $F$. Hence, it is immediately assumed that these coefficients are symmetric in the corresponding indices, such as, for instance, $\beta_{ijk} = \beta_{ikj}$.
A similar reasoning applies to $\chi^{{}_{(2)}}_{ijk}$ and $\tilde\chi^{{}_{(2)}}_{ijk}$ coefficients, but now involving all three of their indexes. 
It should be mentioned, however, that when a dispersive medium is considered additional considerations are required in order to determine the symmetry properties of these coefficients (see, for instance, the discussion in Appendix~\ref{nonlocal}).

The magnetization is here defined to have the same physical dimension of the auxiliary field $H_i$, as follows from the formulation of the Maxwell-Amp\`ere law in terms of the $H_i$ field in Eq.~(\ref{2b}). 
However, the choice of magnetization with the same physical dimension of $B_i $ is also used in the literature \cite{1973-schmid,2005JPhD...38R.123F,2009EPJB...71..299R,2019PhRvA..88841}. 
Finally, all the above discussed optical coefficients are conveniently defined in terms of the derivatives of $F$ with respect to the electric and magnetic fields in the different orders of the expansion. Hence, differences in their notation are common in the literature. 

\section{A note on non instantaneous response of a material medium}
\label{nonlocal}
As is well known, polarization and magnetization phenomena appears as consequence of externally applied fields. Hence, such effects must be causally connected to the applied fields by means of response functions that characterize each particular medium in consideration \cite{post1962}. Here, a brief account of this issue is presented in order to better understand the approximations implemented in this manuscript. Possible effects due to spatial dispersion are being neglected.

We start with the auxiliary field $D_i(\vec r,t) = \varepsilon_0 E_i(\vec r,t) + P_i(\vec r,t)$ (to simplify the notation, the spatial dependence is omitted hereinafter). Keeping only a linear dielectric contribution to the polarization vector, it reads, 
\begin{align}
P_i(t) = \varepsilon_0 \int_{-\infty}^{\infty}d\tau f_{ij}^{{}_{(1)}}(\tau)E_j(t-\tau),
\nonumber
\end{align}
where $f_{ij}^{{}_{(1)}}(\tau)$ is the response function of the medium that obeys $f_{ij}^{{}_{(1)}}(\tau) = 0$, $\tau <0$, so as to ensure causality. As assumed in Sec.~\ref{wave-propagation}, the total field is split in a constant background field plus a monochromatic wave field of frequency $\omega$ as $E_i(t) = E_i^0+E_i ^\omega(t)$, with $E_j^{\omega}(t) = e_j \,\mbox{exp}[i(\omega t-q_n x_n)]$. Now, introducing this result in $P_i (t)$, returning to $D_i(t)$, and organizing it in terms of background and wave fields, we get
\begin{align}
D_j(t) =&\; \varepsilon_0 \left[ \delta_{jk} + \int_{-\infty}^{\infty}d\tau f_{jk}^{{}_{(1)}}(\tau) \right] E^0_k
\nonumber\\
&+\varepsilon_0\left[ \delta_{jk} + \int_{-\infty}^{\infty}d\tau f_{jk}^{{}_{(1)}}(\tau) {\rm e}^{-i\omega\tau} \right] E^\omega_k(t).
\nonumber
\end{align}
The first-order frequency-dependent electric susceptibility of the medium is defined as 
\begin{align}
\chi_{jk}^{{}_{(1)}}(\omega) = \int_{-\infty}^{\infty}d\tau f_{jk}^{{}_{(1)}}(\tau) {\rm e}^{-i\omega\tau}.
\nonumber
\end{align}
Hence, it follows that $D_i(t)$ can be separated into a constant part, that depends on the magnitude of the background field, and a wave part as
\begin{align}
D_i(t) =\varepsilon_0 \left[ \delta_{ij} + \chi_{ij}^{{}_{(1)}}(0) \right] E^0_j
+\varepsilon_0\left[ \delta_{ij} + \chi_{ij}^{{}_{(1)}}(\omega) \right] E^\omega_j(t).
\nonumber
\end{align}
In the study of wave propagation only the time derivative of $D_i(t)$ is required, which makes the first term in the above equation inconsequential.

Similar reasoning applies when nonlinear contributions to electric and magnetic susceptibilities are involved. For instance, if a first-order nonlinear contribution to the dielectric sector (the one responsible by Pockels' effect) is included, we would obtain,
\begin{align}
D_i(t) =& \;\varepsilon_0 \left[ \delta_{ij} + \chi_{ij}^{{}_{(1)}}(0) +  \chi_{ijk}^{{}_{(2)}}(0,0)E^0_k \right] E^0_j 
\nonumber \\
& +\varepsilon_0\left[ \delta_{ij} + \chi_{ij}^{{}_{(1)}}(\omega)+ 2\chi_{ijk}^{{}_{(2)}}(\omega,0)E^0_k \right] E^\omega_j(t)
\nonumber \\
& +\varepsilon_0 \chi_{ijk}^{{}_{(2)}}(\omega,\omega)E^\omega_j(t)E^\omega_k(t),
\nonumber
\end{align}
where the second-order electric susceptibility $\chi_{ijk}^{{}_{(2)}}(\omega,\omega')$ is defined as
\begin{align}
\chi_{jkl}^{{}_{(2)}}(\omega,\omega') = \int_{-\infty}^{\infty}d\tau\int_{-\infty}^{\infty}d\tau' f_{jkl}^{{}_{(2)}}(\tau,\tau') {\rm e}^{-i(\omega\tau +\omega'\tau')},
\nonumber
\end{align}
and $f_{jkl}^{{}_{(2)}}(\tau,\tau')$ is the corresponding response function. 
Intrinsic permutation symmetry is being assumed, which means that  $\chi_{ijk}^{{}_{(2)}}(\omega,\omega')=\chi_{ikj}^{{}_{(2)}}(\omega',\omega)$.
Given the physical regime assumed in the study of wave propagation in Sec.~\ref{wave-propagation}, only the second term in the above expression for $D_i(t)$ collaborates because the first one is a constant and the last one is second-order in the wave field, i.e., it is negligible in our treatment because $\| E_i^0\| \gg \| E_i^\omega(t)\|$. 
Thus, 
\begin{align}
\partial_t D_i(t) 
\approx \varepsilon_0\left[ \delta_{ij} + \chi_{ij}^{{}_{(1)}}(\omega)+ 2\chi_{ijk}^{{}_{(2)}}(\omega,0)E^0_k \right]  \partial_t E^\omega_j \,.
\nonumber
\end{align}
Ignoring the frequency dependence in the susceptibilities for a moment, the above result can be directly obtained by assuming $\varepsilon_{ij} = \varepsilon_0[ \delta_{ij} + \chi_{ij}^{{}_{(1)}}+ \chi_{ijk}^{{}_{(2)}}E_k(t) ]$ in Eq.~(\ref{constitutive1}), taking the time derivative of $D_i(t)$, and implementing the approximations discussed in Sec.~\ref{wave-propagation}. 

Direct inspection of the above equation shows that the nonlinear effect is activated by the constant background electric field $E_i^0$. In the absence of this field, the system behaves like a linear anisotropic system, up to a second-order term in the wave field that was neglected in our approximations. Finally, the idealized case of an instantaneous medium, would produce a similar result. In this case, the response functions would be proportional to delta functions and the above result would be the same, but with electric susceptibilities taking their static values $\chi_{ij}^{{}_{(1)}}(0)$ and $\chi_{ijk}^{{}_{(2)}}(0,0)$. 
\bibliography{../mybibfiles/optics}

\end{document}